\documentclass{article}
\usepackage{waspaa21,amssymb,amsmath,graphicx,times,url,bm}
\usepackage{color,cite}
\usepackage{booktabs,tabularx}
\usepackage{comment}
\usepackage{hyperref}

\usepackage[linesnumbered,ruled,vlined]{algorithm2e}
\SetKwInput{KwInput}{Input}                
\SetKwInput{KwOutput}{Output}              

\DeclareMathOperator{\encoder}{Enc}
\DeclareMathOperator{\decoder}{Dec}

\DeclareMathOperator{\glu}{GLU}
\DeclareMathOperator{\swish}{Swish}

\DeclareMathOperator{\prelu}{PReLU}

\DeclareMathOperator{\dropout}{Dropout}

\DeclareMathOperator{\dense}{Dense}
\DeclareMathOperator{\ffn}{FeedForwardModule}
\DeclareMathOperator{\depthconv}{DepthwiseConv1D}
\DeclareMathOperator{\instNorm}{InstanceNorm}
\DeclareMathOperator{\layerNorm}{LayerNorm}
\DeclareMathOperator{\batchNorm}{BN}
\DeclareMathOperator{\scaling}{Scale}
\DeclareMathOperator{\pow}{pow}

\DeclareMathOperator{\mhsa}{MhsaModule}
\DeclareMathOperator{\favor}{MhsaFavorModule}

\title{DF-Conformer: Integrated architecture of Conv-TasNet and Conformer using linear complexity self-attention for speech enhancement}

\name{
\parbox{0.99\linewidth}{
\centering
Yuma Koizumi,
Shigeki Karita,
Scott Wisdom,
Hakan Erdogan,
John R. Hershey,\\
Llion Jones,
Michiel Bacchiani
}}
\address{Google Research}

\begin{document}

\ninept
\maketitle

\begin{sloppy}

\begin{abstract}
Single-channel speech enhancement (SE) is an important task in speech processing. A widely used framework combines an analysis/synthesis filterbank with a mask prediction network, such as the Conv-TasNet architecture. In such systems, the denoising performance and computational efficiency are mainly affected by the structure of the mask prediction network. In this study, we aim to improve the sequential modeling ability of Conv-TasNet architectures by integrating Conformer layers into a new mask prediction network. To make the model computationally feasible, we extend the Conformer using linear complexity attention and stacked 1-D dilated depthwise convolution layers. We trained the model on 3,396 hours of noisy speech data, and show that (i) the use of linear complexity attention avoids high computational complexity, and (ii) our model achieves higher scale-invariant signal-to-noise ratio than the improved time-dilated convolution network (TDCN++), an extended version of Conv-TasNet.
\end{abstract}


\section{Introduction}
\label{sec:intro}

Speech enhancement (SE) is the task of recovering target speech from a noisy signal~\cite{dlwang_2018}. 
In addition to its applications in telephony and video conferencing~\cite{DnsIcassp2021}, single-channel SE is a basic component in larger systems, such as multi-channel SE~\cite{Erdogan+2016,Higuchi2018}, multi-modal SE~\cite{Gabbay2018,Afouras2018,Gu2020,tzinis2021into}, and automatic speech recognition (ASR)~\cite{hakan_2015,kinoshita_2020,espnetse_2021} systems. Therefore, it is important to improve both the denoising performance and the computational efficiency of single-channel SE.

In recent years, rapid progress has been made on SE using deep neural networks (DNNs)~\cite{dlwang_2018}. Conv-TasNet~\cite{convtasnet} is a powerful model for SE that uses a combination of trainable analysis/synthesis filterbanks~\cite{asteroid} and a mask prediction network using stacked 1-D dilated depthwise convolution (1D-DDC) layers. Since the denoising performance and computational efficiency are mainly affected by the mask prediction network, one of the main research topics in SE is improving the mask prediction architecture~\cite{Kavalerov2019,Scott2020,DPtransformer,sepformer,tstnn2021,LuoIcassp2021,BarunIcassp2021}. For example, the improved time-dilated convolution network (TDCN++)~\cite{Kavalerov2019,Scott2020} extended Conv-TasNet to improve SE performance.

A promising candidate for improving mask prediction networks is the Conformer architecture. The Conformer~\cite{conformer} architecture has been shown to be effective in ASR~\cite{conformer}, diarization~\cite{Soumi2021}, and sound event detection~\cite{Miyazaki2020,Hayashi2020}. Conformer is derived from the Transformer~\cite{transformer} architecture by including 1-D depthwise convolution layers to enable more effective sequential modeling.

In this paper we combine Conformer layers with the dilated convolution layers of the TDCN++ architecture. However, this introduces two critical problems related to the short window and hop sizes used in trainable analysis/synthesis filterbanks. The first problem is large computational cost because the time-complexity of the multi-head-self-attention (MHSA) in the Conformer has a quadratic dependence on sequence length. Secondly, the small hop-size of neighboring time-frames reduces the temporal reach of sequential modeling when using temporal convolution layers.

In order to make the model computationally feasible, we use a linear-complexity variant of self-attention in the Conformer, known as \emph{fast attention via positive orthogonal random features} (FAVOR+), as used in Performer~\cite{performer}. These ideas are partly inspired by the local-global network for speaker diarization using a time-dilated convolution network (TDCN)~\cite{Soumi2021} which shows that the combination of a linear complexity self-attention and a TDCN improves both local and global sequential modeling. We show in experiments below that the resulting model, which we call the \emph{dilated FAVOR Conformer} (DF-Conformer), achieves better enhancement fidelity than the TDCN++ of comparable complexity.

\section{Preliminaries}
\label{sec:related}

\subsection{Conv-TasNet and its extensions on speech enhancement}
\label{sec:conv_tasnet}

Let the $T$-sample time-domain observation $\bm{x} \in \mathbb{R}^T$ be a mixture of a target speech $\bm{s}$ and noise $\bm{n}$ as $\bm{x} = \bm{s} + \bm{n}$, where $\bm{n}$ is assumed to be environmental noise and does not include interference speech signals. The goal of SE is to recover $\bm{s}$ from $\bm{x}$.

In mask-based SE, a mask is estimated using a mask prediction network and applied to the representation of $\bm{x}$ encoded by an encoder, then the estimated signal $\bm{y} \in \mathbb{R}^T$ is re-synthesized using a decoder. The enhancement procedure can be written as
\begin{align}
\bm{y} = \decoder \left(\encoder(\bm{x}) \odot \mathcal{M}(\encoder(\bm{x})) \right)
\end{align}
where 
$\encoder: \mathbb{R}^T \to \mathbb{R}^{N \times D_e}$ and 
$\decoder: \mathbb{R}^{N \times D_e} \to \mathbb{R}^T$
are the signal encoder and decoder, respectively, $D_e$ is the encoder output dimension, $\odot$ is the element-wise multiplication, and $\mathcal{M}: \mathbb{R}^{N \times D_e} \to [0,1]^{N \times D_e}$ is the mask prediction network. Early studies used the short-time-Fourier-transform (STFT) and the inverse-STFT (iSTFT) as encoder and decoder~\cite{hakan_2015,Koizumi2018}, respectively. More recent studies use a trainable encoder/decoder~\cite{convtasnet} which are often called trainable ``filterbanks''~\cite{Pariente2020}, e.g. in Asteroid~~\cite{asteroid}.

One of the main research topic in SE is the design of the network architecture of $\mathcal{M}$, because the performance and computational efficiency of SE are mainly affected by the structure of $\mathcal{M}$. Conv-TasNet~\cite{convtasnet} is a powerful model for speech separation and SE, and whose $\mathcal{M}$ consists of stacked 1D-DDC layers. TDCN++~\cite{Kavalerov2019,Scott2020} is an extension of Conv-TasNet. The main difference of TDCN++ with Conv-TasNet is the use of instance norm instead of global layer norm and the addition of explicit scale parameters after each dense layer. The pseudo-code for $\mathcal{M}$ in the TDCN++ is shown in \textbf{Algorithm \ref{algo:tdcn}}. TDCN++ consists of $L$ stacked TDCN-blocks, and each TDCN-block mainly consists of two dense layers for frame-wise feature modeling and one 1D-DDC layer for sequence modeling. The dilation factor $d$ increases exponentially to ensure a sufficiently large temporal context window to take advantage of the long-range dependencies of the speech signal, and $L_s$ TDCN-blocks are repeated $R$ times where $L = RL_s$. The time complexity of TDCN++ is roughly proportional to $O(LN)$ when $N \gg D_b$, where $D_b$ is the input dimension of TDCN-blocks.

\begin{algorithm}[t]
\DontPrintSemicolon
\caption{$\mathcal{M}$ of TDCN++~\cite{Kavalerov2019,Scott2020} where $\bm{X} = \encoder(\bm{x})$ and $\sigma$ is logistic sigmoid function.}
\label{algo:tdcn}
\SetKwFunction{FSum}{PreprocessDilatedConv}
\SetKwFunction{FDev}{PostprocessDilatedConv}
\SetKwFunction{FMain}{MaskPredictorOfTDCN++}
\SetKwFunction{FSub}{TdcnBlock}
\SetKwProg{Fn}{Function}{:}{\KwRet}
\Fn{\FMain{$\bm{X}$}}{
    $\bm{z} \gets \dense(\bm{X})$ \tcp*{$\mathbb{R}^{N \times D_e}_{+} \to \mathbb{R}^{N \times D_b}$}
    \For{$i = 1$ to $L$}{
        $d \gets \pow(2, \mod(i-1, L_s))$\;
        $\bm{z} \gets \bm{z} + \mbox{$i$-th } \FSub\left(\bm{z}, d\right)$\;
    }
    $\bm{M} \gets \sigma\left(\dense(\bm{z})\right)$ \tcp*{$\mathbb{R}^{N \times D_b} \to [0,1]^{N \times D_e}$}
    \KwRet $\bm{M}$
}
\SetKwProg{Fn}{Function}{:}{}
\Fn{\FSub{$\bm{z}$, $d$}}{
    $\bm{z} \gets \dense(\bm{z})$ \tcp*{$\mathbb{R}^{N \times D_b} \to \mathbb{R}^{N \times D_c}$}
    $\bm{z} \gets \instNorm \left( \prelu \left( \scaling (\bm{z}) \right) \right)$\;
    $\bm{z} \gets \depthconv \left( \bm{z}, \mbox{dilation=}d \right)$\;
    $\bm{z} \gets \instNorm \left( \prelu \left( \bm{z} \right) \right)$\;
    $\bm{z} \gets \dense(\bm{z})$ \tcp*{$\mathbb{R}^{N \times D_c} \to \mathbb{R}^{N \times D_b}$}
    \KwRet $\scaling (\bm{z})$\;
}
\end{algorithm}

\subsection{Conformer}
\label{sec:conformer}

Conformer~\cite{conformer} is a derived model of Transformer~\cite{transformer} that was originally proposed for ASR~\cite{conformer} and later adopted in audio-related applications such as audio event detection~\cite{Miyazaki2020,Hayashi2020} and speech separation~\cite{chen2020continuous}. The structure of the Conformer is similar to the TDCN++, in that it consists of $L$ stacked Conformer-blocks~\cite{conformer}. \textbf{Algorithm \ref{algo:conformer}} shows the pseudo-code of a Conformer-block. By comparing \textbf{Algorithm \ref{algo:tdcn}} and \textbf{\ref{algo:conformer}}, we can see that the constituent layers of the Conformer-block and the TDCN-block are also similar; one Conformer-block mainly consists of several dense layers for frame-wise feature modeling, and one 1-D depthwise convolution layer and one MHSA-module for sequence modeling~\cite{conformer}. One of the main differences between the TDCN-block and the Conformer-block is the MHSA-module. Conformer enables global sequence modeling by using MHSA-modules instead of dilated depthwise convolution layers with local receptive fields.

\begin{algorithm}[t]
\DontPrintSemicolon
\caption{Conformer block~\cite{conformer}. $\batchNorm$ means batch normalization. For details of $\mhsa$ and $\ffn$, see~\cite{conformer}.}
\label{algo:conformer}
\SetKwFunction{FSub}{ConformerBlock}
\SetKwFunction{FSum}{PreprocessConv}
\SetKwFunction{Fmain}{PostprocessConv}
\SetKwProg{Fn}{Function}{:}{}
\Fn{\FSub{$\bm{z}$}}{
    $\bm{z} \gets \bm{z} + \frac{1}{2}\ffn(\bm{z})$\;
    $\bm{z} \gets \bm{z} + \mhsa(\bm{z})$\;
    $\bm{r} \gets \glu \left(\dense \left(\layerNorm(\bm{z}) \right)\right)$\;
    $\bm{r} \gets \depthconv \left( \bm{r}, \mbox{dilation=}1 \right)$\;
    $\bm{z} \gets \bm{z} + \dropout \left( \dense \left(\swish \left(\batchNorm(\bm{r}) \right)\right)\right))$\;
    $\bm{z} \gets \bm{z} + \frac{1}{2}\ffn \left(\bm{z} \right)$\;
    \KwRet $\layerNorm(\bm{z})$\;
}
\end{algorithm}

\section{Proposed method}
\label{sec:propose}

In this section, we first describe two problems for incorporating the Conformer into TDCN++ framework in Sec~\ref{sec:problem}, and our solutions for each problem are described in Sec. \ref{sec:favor} and \ref{sec:dilated_mhsa}, respectively.

\subsection{Model structure and computational challenges}
\label{sec:problem}

Based on the successes of Conformer in speech-related tasks, we aim to replace the TDCN blocks in TDCN++ with Conformer-blocks. Unfortunately, the simple combination of trainable filterbanks and Conformer-blocks causes two critical problems. These problems are caused by the short window size of 2.5 ms and hop size of 1.25 ms used in trainable filterbanks for short-time analysis of the input signal.

\vspace{1pt}
\textbf{Problem 1:} The computational complexity. The computational cost of MHSA-module is quadratic in the number of frames $N$. In the original Conformer model~\cite{conformer}, convolutional subsampling limits the size of $N$. For example, for a 1 second signal, $N$ is 25. In contrast, for TDCN++, the same signal would result in $N=500$.

\vspace{1pt}
\textbf{Problem 2:} The receptive field for sequence modeling is insufficient. The original Conformer has a hop-size of 40 ms, while the standard trainable filterbank has a hop-size of 1.25 ms. This means that the receptive field for depthwise convolution is 6.25 ms when using the default kernel size of 5, which may degrade the accuracy of the analysis of local changes in the signal.

One possible approach is to use the dual-path approach~\cite{DPtransformer,sepformer,tstnn2021}, which is equivalent to using sparse and block diagonal attention matrices corresponding to the inter- and intra-transformers, respectively. Alternatively, we use FAVOR+ attention introduced in Performer~\cite{performer} which has linear computational complexity: $O(N)$. The novelty in our approach comes from using linear FAVOR+ attention to replace softmax-dot-product attention as well as performing local analysis with 1D-DDC to replace non-dilated convolutions in Conformer. Based on these two characteristics of the proposed method, we name our $\mathcal{M}$ as dilated-FAVOR-Conformer (DF-Conformer), and $L$-layer DF-Conformer is referred as DF-Conformer-$L$. The pseudo-code of DF-Conformer-$L$ is shown in \textbf{Algorithm \ref{algo:dfconformer}}. The time complexity of DF-Conformer-$L$ is also roughly in proportion to $O(LN)$ when $N \gg D_b$.

\begin{algorithm}[t]
\DontPrintSemicolon
\caption{$\mathcal{M}$ using DF-Conformer-$L$. Red lines are differences from TDCN++ and Conformer-block.}
\label{algo:dfconformer}
\SetKwFunction{FSub}{DF-ConformerBlock}
\SetKwFunction{FSum}{PreprocessConv}
\SetKwFunction{FDev}{PostprocessConv}
\SetKwFunction{FMain}{DF-Conformer}
\SetKwProg{Fn}{Function}{:}{\KwRet}
\Fn{\FMain{$\bm{X}$}}{
    $\bm{z} \gets \dense(\bm{X})$ \tcp*{$\mathbb{R}^{N \times D_e}_{+} \to \mathbb{R}^{N \times D_b}$}
    \For{$i = 1$ to $L$}{
        $d \gets \pow(2, \mod(i-1, L_s))$\;
        \textcolor{red}{$\bm{z} \gets \bm{z} + \mbox{$i$-th } \FSub\left(\bm{z}, d\right)$}\;
    }
    $\bm{M} \gets \sigma\left(\dense(\bm{z})\right)$ \tcp*{$\mathbb{R}^{N \times D_b} \to [0,1]^{N \times D_e}$}
    \KwRet $\bm{M}$
}
\SetKwProg{Fn}{Function}{:}{}
\Fn{\FSub{$\bm{z}$, \textcolor{red}{$d$}}}{
    $\bm{z} \gets \bm{z} + \frac{1}{2}\ffn(\bm{z})$\;
    \textcolor{red}{$\bm{z} \gets \bm{z} + \favor(\bm{z})$}\;
    $\bm{r} \gets \glu \left(\dense \left(\layerNorm(\bm{z}) \right)\right)$\;
    \textcolor{red}{$\bm{r} \gets \depthconv \left( \bm{r}, \mbox{dilation=}d \right)$}\;
    $\bm{z} \gets \bm{z} + \dropout \left( \dense \left(\swish \left(\batchNorm(\bm{r}) \right)\right)\right))$\;
    $\bm{z} \gets \bm{z} + \frac{1}{2}\ffn \left(\bm{z} \right)$\;
    \KwRet $\layerNorm(\bm{z})$\;
}
\end{algorithm}

\subsection{Linear time-complexity MHSA-module using FAVOR+}
\label{sec:favor}

Recently, many extended Transformer architectures have been proposed to make improvements around computational and memory efficiency~\cite{tay2020efficient, katharopoulos2020transformers}. Performer~\cite{performer} is one of them; it is an $O(N)$ Transformer architecture which uses FAVOR+. 
In self-attention, the query, $\mathbf{Q}$, key, $\mathbf{K}$, and value, $\mathbf{V}$ $\in \mathbb{R}^{N \times D}$ are combined as 
$\operatorname{sa}(\mathbf{Q},\mathbf{K},\mathbf{V})=\operatorname{softmax}(\mathbf{Q}\mathbf{K}^{\top})\mathbf{V}$.
In FAVOR+, this is approximated as $\operatorname{sa}(\mathbf{Q},\mathbf{K},\mathbf{V})\approx \mathbf{D}^{-1}\phi(\mathbf{Q})\big(\phi(\mathbf{K})^{\top}\mathbf{V}\big)$, for a suitable feature map $\phi(\cdot)$ applied to the rows of each matrix, avoiding the quadratic term $\mathbf{Q}\mathbf{K}^{\top}$. Here $\mathbf{D}$ is a normalizing diagonal matrix with $\operatorname{diag}(\mathbf{D})=\phi(\mathbf{Q})(\phi(\mathbf{K})^{\top} \mathbf{1})$, and $\mathbf{1} \in \mathbb{R}^{N}$ an all ones vector. This approximation is made accurate in FAVOR+ using a random projection based non-negative valued $\phi(\cdot)$ of a suitable size~\cite{performer}. To implement this idea, we replace the softmax-dot-product self-attention in \textbf{Algorithm \ref{algo:conformer}} with FAVOR+ self-attention. Hereafter, we refer to this new module as ``MHSA-FAVOR-module''.

\subsection{Use of dilated depthwise convolution in Conformer}
\label{sec:dilated_mhsa}

We strengthen the network's temporal analysis capability by using 1D-DDC instead of the standard 1-D depthwise convolution used in the Conformer-blocks. As in TDCN++, we use an exponentially increasing dilation factor $d$. To implement this idea, DF-Conformer-block also takes $d$ as an argument, and it is passed to the 1D-DDC layer as the dilation parameter.

In a similar strategy as~\cite{Soumi2021} and DF-Conformer, MHSA-FAVOR-module can also be incorporated into the TDCN-block. As an alternative network architecture, we insert $\bm{z} \gets \bm{z} + \favor(\bm{z})$ between line 9 and 10 of \textbf{Algorithm \ref{algo:tdcn}}, and refer to it as ``Conv-Tasformer''.

\section{Experiments}
\label{sec:experiment}

We conducted ablation studies and objective experiments in Sec.\,\ref{sec:verif_eval} and \ref{sec:objective_eval}, respectively. Audio demos are available\footnote{\href{https://google.github.io/df-conformer/waspaa2021/}{\texttt{google.github.io/df-conformer/waspaa2021/}}}.

\subsection{Experimental setup}
\label{sec:exp_setup}

\textbf{Dataset: }We used the same dataset used in the SE experiment of~\cite{Scott2020}.
This dataset uses speech from LibriVox
(\href{https://librivox.org/}{\texttt{librivox.org}})
and non-speech sounds from 
\href{https://freesound.org/}{\texttt{freesound.org}}
. The duration of all samples were 3 sec, and sampling rate was 16 kHz. Training, validation, and test datasets consisted of 4,076,102 (3396.8 hours), 7,417 (6.2 hours), and 7,387 (6.2 hours) examples, respectively.
We mixed speech and noise samples in the same manner of~\cite{ScottIcassp2020}. The minimum and maximum signal-to-noise ratio (SNR) of noisy input were $-40$ dB and $45$ dB, respectively, and the average extended short-time objective intelligibility measure (ESTOI)~\cite{estoi} score was 63.7\%.

\vspace{2pt}
\noindent
\textbf{Loss function: }
We estimated masks for both speech and noise in the same manner of~\cite{Scott2020,kinoshita_2020}. Each mask was multiplied with $\encoder(\bm{x})$ and re-synthesized to the time-domain using the same decoder. A mixture consistency projection layer~\cite{ScottIcassp2020} was applied to ensure the mixture of estimated speech and noise equals the noisy input. Finally, the negative thresholded SNR~\cite{Scott2020} loss\footnote{
$\mathcal{L} = -10 \log_{10}(\lVert \bm{s}\rVert^2 / (\lVert \bm{s} -  \bm{y}\rVert^2 + \tau\lVert \bm{s}\rVert^2))$ where $\tau = 10^{\alpha / 10}$ a soft threshold that clamps the loss at $\alpha$ dB. In this study, we used $\alpha = 30$.}
was calculated for both speech and noise, and mixed by weighting 0.8 for speech and 0.2 for noise.

\vspace{2pt}
\noindent
\textbf{Comparison of methods and hyper-parameters: }
For the ablation studies in Sec.\,\ref{sec:verif_eval}, we used three Conformer-based models. The first model is Conformer-$L$ which simply replaces TDCN-blocks in TDCN++ with Conformer-blocks. The second model is F-Conformer-$L$ which is a model that uses only FAVOR+ in DF-Conformer-$L$. The last model is Conformer-$L$-STFT which uses STFT and iSTFT as $\encoder$ and $\decoder$, respectively. For Conformer-$L$-STFT models, we estimated a complex-valued mask~\cite{KoizumiIcassp2020}. We cannot increase the number of parameters of Conformer-$L$ due to its computational complexity, therefore, we used two different model sizes; 3.7M and 8.75M parameters. The former size was determined according to the maximum model size of Conformer-$L$ that can be trained on third-generation Tensor Processing Units (TPUv3). The latter size is that of TDCN++ used in previous studies~\cite{Kavalerov2019,Scott2020}. The hyper parameters were $L=4$ and $D_b=192$ were used for 3.7M models, and $L=8$, $L_s=4$, and $D_b=216$ were used for 8.75M models. For both model sizes, $6$ attention heads and $D_r = 384$ random projection features were used in FAVOR+.

For the SE performance evaluation in Sec.\,\ref{sec:objective_eval}, we compared DF-Conformer-$L$ and Conv-Tasformer with TDCN++~\cite{Kavalerov2019,Scott2020} to confirm the superiority of the proposed models from its base model. In TDCN++, we used the same setting used in~\cite{ScottIcassp2020}, namely, $L=32$, $L_s =8$, $D_b=256$ and $D_c=512$. In Conv-Tasformer, we used the same setting of TDCN++ except $L=16$ and $D_r=128$ to reduce the number of parameters.

For all models, $D_e=256$, and the window and hop sizes of trainable filterbanks were 2.5 ms and 1.25 ms, respectively. For STFT,
the window and hop sizes were 30 ms and 10 ms, respectively, and fast-Fourier-transform size was 512.
All models were trained for 500k steps on 128 Google TPUv3 cores with a global batch size of 512.
We configured the Adam optimizer~\cite{kingma2014adam} with weight decay 1e-6, and learning rate schedule~\cite{transformer} of $D_b^{-0.5} \min(n \times 25000^{-1.5}, n^{-0.5})$, where $n$ is a number of training steps. We clipped the gradient by global $\ell_2$ norm to 5.0. We stored a separate checkpoint of exponential-moving-averaged weights accumulated over training steps with decay rate 0.9999.

\subsection{Evaluation of FAVOR+}
\label{sec:verif_eval}

\begin{figure}[t]
  \centering
\includegraphics[width=\linewidth,clip]{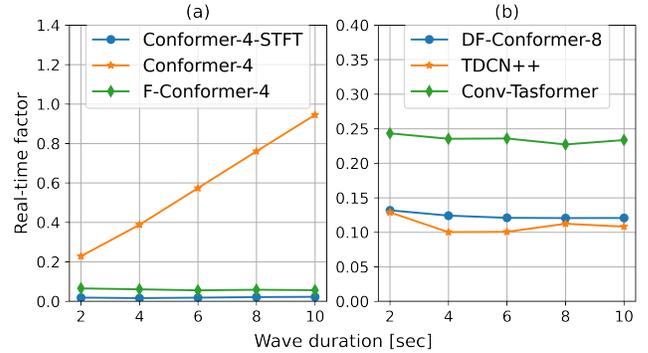} 
  \vspace{-20pt}
  \caption{Comparison of RTF. (a) RTF of Conformer-4 increases as duration of input waveform increases, whereas that of F-Conformer-4 becomes constant. (b) RTFs of DF-Conformer-8 and TDCN++ are comparable, whereas that of Conv-Tasformer is larger than others due to additional MHSA-FAVOR-block.}
  \label{fig:processing_time}
\end{figure}

To confirm the effects of FAVOR+, we compared the real-time factor (RTF) of Conformer-4-STFT, Conformer-4, and F-Conformer-4 using 1 CPU. Figure\,\ref{fig:processing_time}\,(a) shows the comparison results. In the case of Conformer-4-STFT, RTF does not increase significantly because $N$ was $100/\mbox{sec}$ in our STFT setting and it is still feasible with $O(N^2)$ MHSA-module. Whereas RTF of Conformer-4 increases linearly as $N$ was $500/\mbox{sec}$ in our trainable filterbank setting and MHSA-module. Since the time-complexity of FAVOR+ is in proportion to $O(N)$, F-Conformer-4 has solved this problem.

\begin{table}[ttt]
\caption{Results of evaluation for FAVOR+. Prefix ``F'' means the use of FAVOR+, and postfix ``STFT'' means the use of STFT and iSTFT for $\encoder$ and $\decoder$, respectively.}
\label{tab:exp_favor}
\centering
\begin{tabular}{ c| c | ccc }
\toprule
\textbf{Model} 	& \textbf{\#Params} 	& \textbf{SI-SNRi}	& \textbf{ESTOI}	& \textbf{RTF}\\	
\midrule
Conformer-4-STFT       &   3.82 M   & 12.47 & 83.4 & \textbf{0.02} \\
Conformer-4       &   3.74 M   & \textbf{13.91} & \textbf{84.8} & 0.31 \\
F-Conformer-4       &   3.59 M   & 12.40 & 80.5 & 0.06 \\	
\midrule
Conformer-8-STFT       &   9.30 M   & 12.64 & \textbf{84.5} & \textbf{0.03} \\
F-Conformer-8       &   8.83 M   & \textbf{13.81} & 83.7 & 0.13 \\
\bottomrule
\end{tabular}
\end{table}

We also compared these methods using two objective metrics; scale-invariant SNR improvement (SI-SNRi)~\cite{sisnr} and the ESTOI. Table\,\ref{tab:exp_favor} shows the results. By comparing Conformer-4-STFT and Conformer-4, the use of a trainable filterbak achieved higher scores than STFT as similar to previous studies~\cite{Kavalerov2019}. When using the small-size model, the SI-SNRi score of F-Conformer-4 was almost the same as those on the Conformer-4-STFT. Meanwhile, with the 8.75M models, SI-SNRi of F-Conformer-8 was 1.2 dB higher than that of Conformer-8-STFT, and ESTOI scores of those were almost comparable. These results suggest that the use of FAVOR+ can achieve high time-domain SE performance with a larger model while avoiding the increase in computational complexity.

\subsection{Objective evaluation}
\label{sec:objective_eval}

\begin{table}[ttt]
\caption{Experimental results. Meaning of prefix and postfix are the same as Table\,\ref{tab:exp_favor}. Additional prefixes ``D'' and ``i'' mean the use of 1D-DDC and iterative model, respectively.}
\label{tab:exp_dilated}
\centering
\begin{tabular}{ c| c | ccc }
\toprule
\textbf{Model} 	& \textbf{\#Params} 	& \textbf{SI-SNRi}	& \textbf{ESTOI}	& \textbf{RTF}\\	
\midrule
TDCN++~\cite{Scott2020}       &   8.75 M   & 14.10 & \textbf{85.7} & \textbf{0.10} \\
Conv-Tasformer       &   8.71 M   & 14.36 & 85.6 & 0.25 \\
DF-Conformer-8     &   8.83 M   & \textbf{14.43} & 85.4 & 0.13 \\
\midrule
iTDCN++~\cite{Scott2020}       &   17.6 M   & 14.84 & 87.1 & \textbf{0.22} \\
iConv-Tasformer       &   17.5 M   & 15.25 & \textbf{87.2} & 0.48 \\
iDF-Conformer-8   &   17.8 M   & \textbf{15.28} & 87.1 & 0.26 \\
\midrule
iDF-Conformer-12   &   37.0 M   & 15.93 & 88.4 & 0.46 \\ 
\bottomrule
\end{tabular}
\end{table}

We compared DF-Conformer-8, TDCN++, and Conv-Tasformer using SI-SNRi, ESTOI, and RTF. From the comparison results shown in Table\,\ref{tab:exp_dilated}, DF-Conformer-8 and Conv-Tasformer achieved comparable scores, and these scores were higher than that of TDCN++. Also, by comparing DF-Conformer-8 and F-Conformer-8 in Table\,\ref{tab:exp_favor}, the use of 1D-DDC significantly improved the scores while avoiding to increase RTF. These results suggest that the use of both 1D-DDC and FAVOR+ is effective in SE. We also compared RTF of these methods as shown in Fig.\,\ref{fig:processing_time}\,(b). RTFs of DF-Conformer-8 and TDCN++ were comparable, whereas that of Conv-Tasformer was larger than others due to additional MHSA-FAVOR-block. Therefore, when inserting FAVOR+ in TDCN-block as Conv-Tasformer, it will be necessary to devise the position and number of MHSA-FAVOR-module in order to improve the computational efficiency.

We also compared the iterative extension of these models~\cite{Kavalerov2019}. Using iterative model improved the scores of all methods, and the results tended to be similar to the non-iterative models. Furthermore, we evaluated a larger model as iDF-Conformer-12 with $L=12$, $D_b = 256$, and the number of attention heads were 8. The size of model was determined so that RTF becomes comparable with iConv-Tasformer. As we can see the results, the scores clearly improved using a large model, thus DF-Conformer would be able to scale the performance according to the model size.

\begin{figure}[t]
  \centering
\includegraphics[width=\linewidth,clip]{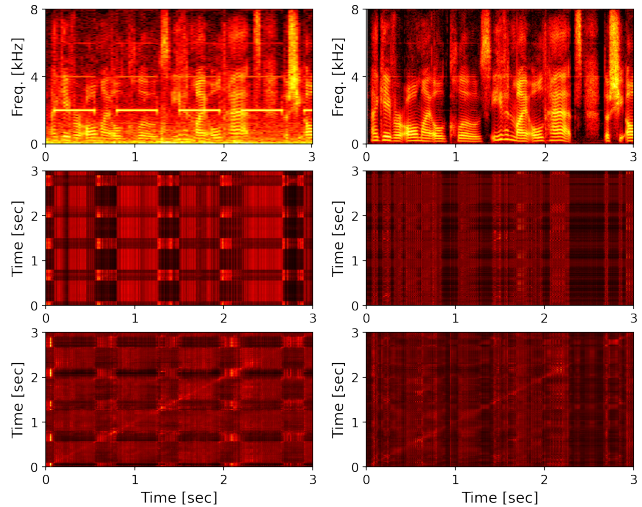} 
  \vspace{-20pt}
  \caption{
  Examples of attention matrices in DF-Conformer-8. 
  Spectrograms of noisy input and enhanced output (top row), and attention matrices for first and third (middle row) and last (bottom row) Conformer blocks calculated by $\mathbf{D}^{-1}\phi(\mathbf{Q})\phi(\mathbf{K})^{\top}$. The x and y axes of attention matrices denote the key and query, respectively.
  }
  \label{fig:attention}
\end{figure}

We finally point out three characteristics in DF-Conformer's attention matrices. First, none of all attention matrices has a local structure that focuses only on nearby time-frames. Secondly, most attention matrices in earlier layers referred to low SNR time-frames to capture the noise characteristics (e.g. Fig.\,\ref{fig:attention} middle-left), or referred to time-frames with similar spectral structures (e.g. Fig.\,\ref{fig:attention} middle-right). Thirdly, some attention matrices of deep layers resemble a sum of a nearly-diagonal matrix and a block matrix (e.g. Fig.\,\ref{fig:attention} bottom). This results suggest that the earlier layers roughly analyze the speech and noise from the entire utterance, and later layers refine the mask based on the local structure.

\section{Conclusion}
\label{sec:conclusion}

In this study, we proposed DF-Conformer which is a Conformer-based time-domain SE network. To improve the computation complexity and local sequential modeling, we extended Conformer using a linear complexity attention mechanism and 1-D dilated separable convolutions. Experimental results showed that (i) the use of a linear complexity attention solves the computational-complexity problems, and (ii) our model achieve higher performance than TDCN++. From the results of experiments, we conclude that DF-Conformer is an effective model for SE. Future works include joint-training of SE and ASR using an all Conformer model, and comparison with the dual-path methods~\cite{DPtransformer,sepformer,tstnn2021} on the SE task.

\setlength{\itemsep}{-0.0pt}
\setlength{\baselineskip}{10.0pt}
\bibliographystyle{IEEEtran}
\footnotesize{
\bibliography{refs21}
}
\end{sloppy}
\end{document}